\documentclass[aip,jcp,amsmath,amsymb,reprint]{revtex4-1}
\usepackage{graphicx}
\usepackage{bm}

\begin{document}

	\title{Designing Pairwise Interactions that Stabilize Open Crystals: Truncated Square and Truncated Hexagonal Lattices}
	\author{William D. Pi\~{n}eros} 
	\affiliation{Department of Chemistry and Biochemistry, University of Texas at Austin, Austin, TX 78712 USA}
	\author{Thomas M. Truskett}
	\affiliation{McKetta Department of Chemical Engineering, University of Texas at Austin, Austin, TX 78712 USA}
	\date{\today}
	\begin{abstract}
		Using a recently introduced formulation of the ground-state inverse design problem for a targeted lattice [Pi\~{n}eros et al., {\emph{ J. Chem. Phys.}} {\bf 144}, 084502 (2016)], we discover purely repulsive and isotropic pair interactions that stabilize low-density truncated square and truncated hexagonal crystals, as well as promote their assembly in Monte Carlo simulations upon isochoric cooling from a high-temperature fluid phase. The results illustrate that the primary challenge to stabilizing very open two-dimensional lattices is to design interactions that can favor the target structure over competing stripe microphases. 
	\end{abstract}

	\maketitle 

\section{Introduction}
	Manufacture of materials with precisely defined nanometer scale structural features remains a formidable challenge. While some top-down fabrication methods (e.g. lithography) have improved significantly in recent years to help address this challenge,\cite{PhotonicMatsDesign,PhotonicMatsDesign2} such approaches remain prohibitively slow and expensive for many commercial applications. Self-assembly--the spontaneous ordering of a material's constituent building blocks to arrive at a targeted equilibrium state--provides a promising (if still nascent) bottom-up alternative to create such nanostructured materials. In self assembly, specific structural control is achieved through systematic modification of the relevant interactions for the building blocks to drive their organization into desired morphologies,\cite{JanusParticlesSelfAssemblyRev,SelfAssemblyforcesReview,SelfAssemblySuperlatticeCatalysis} a strategy enabled through statistical mechanical modeling and recent advances in colloid science and materials chemistry.\cite{ColloidInteractionTuning_1, colloidInteractionsReview,SelfAssemblySuperlatticeCatalysis,SelfAssemblyPatchyParticles1,SelfAssemblyPolyhedraParticles}

	In designing interactions for targeted self assembly, one can consider forward or inverse approaches.\cite{InvDesignTechRev,InvDesignGeneral,InvDesignPerspective} Forward methods often discover new systems via trial and error searches through parameter space, where the key properties of the candidate materials are measured and ranked in terms of their `fitness' relative to those of the target. Such Edisonian approaches, although simple to implement, can unfortunately be inefficient and expensive design strategies. Inverse approaches, on the other hand, offer a more direct means for design, typically via the use of statistical mechanical models solved via constrained optimization algorithms. Though they present significant theoretical and computational challenges, inverse methods can be highly effective at helping to navigate the rugged and high dimensional fitness hypersurfaces encountered in materials design problems.
	
	A classic example of an inverse design problem is the determination of parameters $\{\alpha\}$ of a given isotropic pair potential $\phi(r;\{\alpha\})$ that maximize stability of a specified periodic lattice structure in the ground state. Studies focusing on this type of optimization problem have employed a number of different constraints on the pair potential as well as different objective functions quantifying various aspects of target structure stability, and have consequently discovered a diverse array of interaction types capable of stabilizing even relatively open two-dimensional (2D) and three-dimensional (3D) morphologies (e.g., honeycomb \cite{AvniDimTransfer,MT_SquareHoneyConvexCom}, kagome\cite{InvDesignKagomeDiamond,InvDesignKagome,InvDesignKagomeFunctionalMethod, ZT_MuOptKagomeAsymLats}, simple cubic\cite{Avni3DLattices}, and diamond \cite{Avni3DLattices,InvDesignKagomeDiamond} lattices, to mention a few). For many such cases, systems of particles interacting via the designed pair potentials have been found to spontaneously self-assemble into the target structures upon cooling from a high-temperature fluid phase in Monte Carlo simulations. 

	Ground-state inverse design problems such as these can be formulated as analytical non-linear programs\cite{SquLat_dmu_opt} amenable to high-performance numerical solvers, such as those integrated into GAMS (General Algebraic Modeling System)\cite{GAMSWorldBank,GamsGuide2013,GamsSoftware2013}. This allows for extensive study of theoretical material design questions that were previously inconvenient (or, in cases, impractical) to address with slower converging stochastic optimization methods such as simulated annealing. One recent example relates to understanding qualitative differences between interactions designed to maximize the density range over which a target lattice is the stable ground state versus those designed to maximize the target structure's thermal stability (encoded in the magnitude of the free energy difference between the target and its competitors).\cite{SquLat_dmu_opt} Another pertains to the ability of isotropic pair potentials to stabilize lattices with highly asymmetric angular distributions of particles at a given distance;\cite{KagSnub_gams_opt} an archetypal example of which is the 2D snub-square lattice, which has only a single particle in its third coordination shell.
	
	Here, we adopt this type of formulation to test the extent to which isotropic, repulsive pair potentials can be designed to stabilize ground states of particles organized in low-density periodic lattice structures. We further use Monte Carlo simulations to study whether particles interacting via the designed pair potentials can readily assemble into the target structures from the fluid following a rapid temperature quench. Porous materials such as these, more commonly stabilized by directional attractive interactions (e.g., physical `bonds' between patchy colloids\cite{C4SM00587B,0953-8984-23-40-404206}), can find application in optical\cite{PhotonicMatsDesign}, chemical storage\cite{PorousMat_storage_review,NanoporousMaterials_book}, and separation \cite{NanoporousMaterials_book} technologies. Thus, the discovery of new ways to assemble them from a wide variety of material building blocks and interaction types remains an active area of research.  
	The specific periodic structures that we focus on in this investigation are the 2D truncated square (TS) and truncated hexagonal (TH) lattices, which are characterized by central octagonal or dodecagonal motifs, respectively, that resemble `pores' of empty space within the matrix of surrounding lattice particles. The TH lattice exhibits one of the lowest packing fractions for a 2D close-packed system ($\eta \approx 0.39$) which is approximately half that of the close-packed square lattice and two thirds that of the close-packed honeycomb lattice; the packing fraction of the TS lattice is approximately $12\%$ lower than that of the honeycomb lattice if the two are compared in their respective close-packed states.   
		
	The balance of this article is organized as follows. In section \ref{sec:Methods}, brief descriptions of the ground-state inverse design problem, the strategy we adopt for determining which structures closely compete with the target lattice, and the Monte Carlo simulations that we use to observe assembly from the fluid phase are presented. The results of our study, including the optimized pair potentials and an analysis of the target structures assembled in Monte Carlo simulations, are discussed in section~\ref{sec:Results}, where the differences in designed potentials and assembly behaviors of the TS and TH target lattices are also explored. Concluding remarks and implications of the work are presented in section \ref{Conclusion}. 
\section{Methods} 
\label{sec:Methods}

\subsection{Design Model}
Our design model is framed around an analytical formulation of the inverse ground state problem for a target lattice in terms of constraints on the interparticle interactions [provided by form of the pair potential, $\phi(r;\{\alpha\})$] and an objective function choice. For this work, we define $\phi(r;\{\alpha\})$ as  
 	\begin{equation}
		\begin{aligned}
		\phi(r/\sigma) &= \epsilon \lbrace A(r/\sigma)^{-n} + \sum_{i=1}^{N_h} \lambda_i(1-\tanh[k_i(r/\sigma-d_i)]) \\ 
			       &+f_{\text{shift}}(r/\sigma) \rbrace H[(r_{\text{c}}-r)/\sigma]
		\label{eq:potform}
		\end{aligned}
	\end{equation}
where $A$, $n$,$\lambda_i$, $k_i$, $d_i$ are design parameters (i.e. $\{\alpha\}$ ), $N_h$ is the number of hyperbolic tangent terms used in the pair potential, $H$ is the Heaviside function, $r_{\text{c}}$ is the cut off radius, and $f_{\text{shift}}(r/\sigma)= P (r/\sigma)^2 + Q r/\sigma + R$ is a quadratic shift function added to enforce $\phi(r_{\text{c}}/\sigma)= \phi'(r_{\text{c}}/\sigma)= \phi''(r_{\text{c}}/\sigma)= 0$. In what follows, $N_h=2,3$ for the TS and TH lattice, respectively.
We require $\phi(r/\sigma) >0$ and $\phi'(r/\sigma)<0$ to ensure a monotonically decreasing (i.e., purely repulsive) pair potential which is flexible and can mimic the various soft-repulsive effective (i.e., center-of-mass) interactions that can be observed between, e.g., solvated star polymers, dendrimers, micelles, microgel particles, etc. Of course, additional (or simply different) constraints could be explored in future studies for designing assemblies of specific material systems. For notational convenience, we implicitly nondimensionalize quantities by appropriate combinations of $\epsilon$ and $\sigma$.

As described in detail previously,\cite{SquLat_dmu_opt,KagSnub_gams_opt} with interactions of this type, one can analytically formulate a nonlinear program whose numerical solution provides pair potential parameters that minimize the objective function $F = \sum_j (\mu_t-\mu_{l,j})$.  Here, $\mu_t$ is the zero-temperature [$T=0$] chemical potential of the target lattice at a specified density $\rho_0$, and $\mu_{l,j}$ is that of an equi-pressure lattice $j$ from a specified set of competitive `flag-point' structures (discussed below); the sum is over all such flag-point competitors. In this work, we search for parameters that stabilize the target structure ground state over the widest range of density $\Delta \rho$, while ensuring a chemical potential advantage of the target relative to each flag-point competitor that is greater than a minimum specified threshold (here, we use  $\mu_t-\mu_{l,j} \le -0.01$). Specific information on the program formulation, including the equations used and their numerical solution using solvers in GAMS, is provided elsewhere.\cite{SquLat_dmu_opt,KagSnub_gams_opt}

\subsection{Competing Pool Selection}
    To use the strategy discussed above for designing a pair potential $\phi(r;\{\alpha\})$ that stabilizes a given target structure in the ground state, one first needs to establish a finite (preferably small) pool of the most competitive alternative structures at zero temperature and the same pressure. To do this, we adopt an iterative procedure. First, we carry out a preliminary optimization comparing the chemical potential of the target to others in an initial pool comprising a few select lattice and mesophase structures (e.g., stripes) known to be competitive for systems with isotropic, repulsive interactions.\cite{SquLat_dmu_opt,KagSnub_gams_opt} We then carry out a `forward' calculation that considers more comprehensively equi-pressure competitors. For classes of competing structures that contain free parameters, the values of those parameters are determined by minimizing the chemical potential (using GAMS) under the optimized pair potential (for details see appendix of ref. 25). Any structures that are revealed by this calculation to be more stable than the target lattice are added to the competing pool to be used in the next iteration of the pair potential optimization. This process is repeated until no new structures that closely compete with the target are found in the ground-state phase diagram calculation of the optimized potential.  
	\begin{figure*}[ht]
	\includegraphics[scale=0.20]{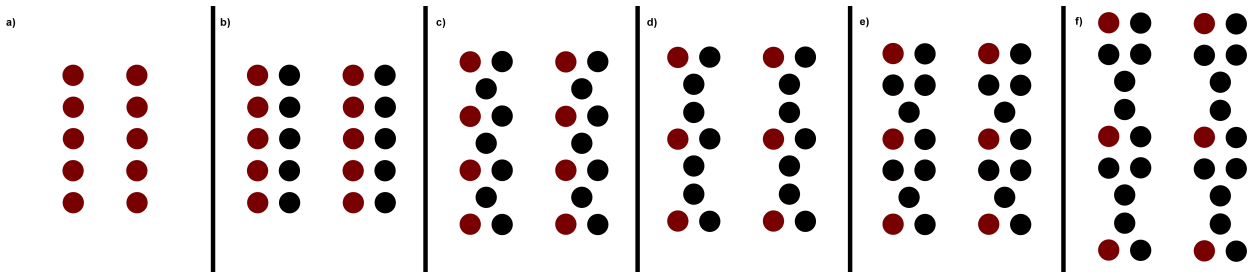}
	\caption{ Competitor stripe classes schematic (a-f). Red particles denote repeating lattice cell and black particles any additional basis. Implicit in each class are numerous possible degrees of freedom, including inter-stripe distance, shears along stripe axis, as well as motif rotations and distortions. Together, these stripes with numerous internal degrees of freedom could be said to represent microphase competitors. }
	\label{fig:competitors}
	\end{figure*} 

 	Unlike for previous ground-state optimizations targeting denser structures,\cite{Avni3DLattices,AvniDimTransfer,SquLat_dmu_opt,KagSnub_gams_opt} the structures within the competing pools for the low-density TS and TH lattices are too numerous to list in detail (totaling 60+).  Instead, it is more insightful to consider competitors as general classes of stripe motifs with a variety of internal degrees of freedom.  This is shown more clearly in schematic figure \ref{fig:competitors} where competitor classes are illustrated in each panel (a-f) and red particles represent fundamental lattice cells.  For example, panel a) shows two stripes of particles separated by a given distance. Possible degrees of freedom include this separation distance as well as shears along the stripe axis, which in this case produce rectangular or oblique lattices. Panels b-f denote similar stripe-like classes, but now with increasing number of particles per cell (black particles) and more specific motifs. Relevant degrees of freedom here include the distance between stripes, shears along the stripe axis, but also more specific possibilities (e.g., motif distortions or rotations). Altogether, these six classes represent stripe microphases that constitute most of the strong competitors found for both design targets of this study. In what follows, we list the final competitor pools for each target as a tally of competitors belonging to each class as well as any general or specialized competitor not included in this set.  

	For the TS lattice target, the final pool of competitors included the following standard periodic lattices that are {\em not} part of the aforementioned stripe classes: square, hexagonal\cite{BravaisNote}, honeycomb, snub square, snub trihexagonal, and distorted kagome (2 competitors). The `stripe-class' competitors for the TS lattice included four structures from class a), three from class b), five from class c), four from class d), and one from class e).  For the TH target, the non-stripe class competitors included the following standard lattices: square, hexagonal, honeycomb, snub trihexagonal, TS, and snub square with aspect ratio $b/a=1.8$. The stripe-class competitors for the TH lattice included seven structures from class a), seven from class b), seven from class c), five from class d), and seven from class f). Additionally, two specialized competitors arose for the TH lattice; one was a cluster of five particles repeating across an open oblique lattice (figure S1) and another was an open decagonal motif with a particle in the center (figure S2). 

	Finally, note that while all competitors were ensured to have chemical potentials greater than those of the target with the optimized interactions, only representative members of each stripe class and other lattices (so-called `flag-point' lattices\cite{SquLat_dmu_opt}) can be effectively used in objective function evaluations for this formulation and in ensuring the minimum required chemical potential advantage of the target described above. For these targets, the particular identity of a stripe class flag-point competitor is not too important as long as the overall flag-point set spans one member of each class. On the other hand, standard lattices (e.g. hexagonal) or uniquely specialized competitors like the decagonal motif structure for TH or snub trihexagonal for TS enter directly as flag-point competitors by default. 

	\begin{figure*}[ht]
	\includegraphics[scale=0.3]{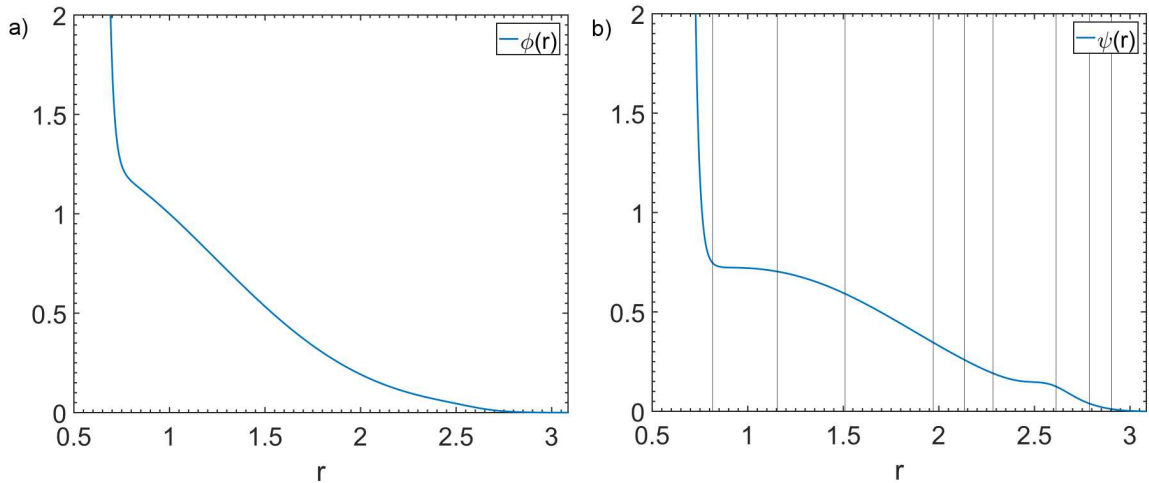}
	\caption{a) Repulsive pair potential $\phi(r)$ designed to stabilize the TS lattice as the ground-state structure, and (b) $\psi(r)$ obtained from $\phi(r)$ via eq.~\ref{eq:psir}. Black vertical lines indicate positions of the first nine coordination shells of the TS crystal at the midpoint of its stable density range ($\rho=1.03$). The parameters of the optimized pair potential are presented in table S1.} 
	\label{fig:truncsqu_potpsi}
	\end{figure*} 
\subsection{Monte Carlo Simulations}
\label{sec:monte_carlo}
    To explore the feasibility of self assembly from fluids of particles interacting via the optimized pair potentials, Monte Carlo simulations were carried out in the canonical ensemble as follows. For the potential optimized for the TS lattice, a system of $N=100$ particles (in a periodically replicated simulation cell with dimensions chosen to fix the number density, $\rho_0=1.03$) was isochorically heated to a high temperature, melting the perfect crystal to form a fluid. The fluid was then isochorically quenched from high temperature back to a crystal at $T=0.0091$. The crystal was then further cooled to $T=0.005$ for structure refinement and computation of the radial distribution function.  For the potential optimized for the TH lattice, a system of $N=96$ particles (in a periodically replicated cell with dimensions set to fix $\rho_0=1.075$) was melted from the perfect crystal to form a fluid. 
    Two dozen identical fluid configurations were seeded with a small frozen crystal of 21 particles pinned into perfect lattice positions. These configurations were then quenched from high temperature to $T=0.06$ over 4 million Monte Carlo steps. For systems displaying assembly of the target structure, the seed particles were subsequently unpinned, and the whole system was allowed to relax for $90,000$ Monte Carlo steps for computation of the radial distribution function. 

\section{Results and Discussion}
\label{sec:Results}
	Using the problem formulation described in section \ref{sec:Methods}, we were able to solve for parameters of the monotonically decreasing pair potential $\phi(r)$ (given by eq. \ref{eq:potform}) that maximize the density range over which the TS lattice is the stable ground state structure (here, $0.98 \leq \rho \leq 1.08$), while also ensuring that the ground state exhibits, at $\rho_0=1.03$, a chemical potential advantage of at least $\Delta\mu=0.01$ over equi-pressure flag-point competitors. Importantly, the latter ensures a significant free energy separation of the target from various closely competing stripe microphases. The resulting pair potential $\phi(r)$ is shown in figure~\ref{fig:truncsqu_potpsi}a, and the list of optimized potential parameters is provided in table S1. As can be seen, $\phi(r)$ has a simple, ramp-like form with a steeply repulsive core at $r\sim 0.7$. This is interesting because particles interacting via a similar hard-core plus linear-ramp repulsion are known to exhibit rich ground-state behavior as a function of density and the parameters of the pair potential,\cite{PhysRevE.58.1478} displaying a variety of periodic crystalline structures (including some with nonequivalent lattice sites or multiple particles per unit cell) as well as a random quasicrystal. 
	\begin{figure}[ht]
	\includegraphics[trim=10cm 1cm 1cm 1cm,clip,scale=0.50]{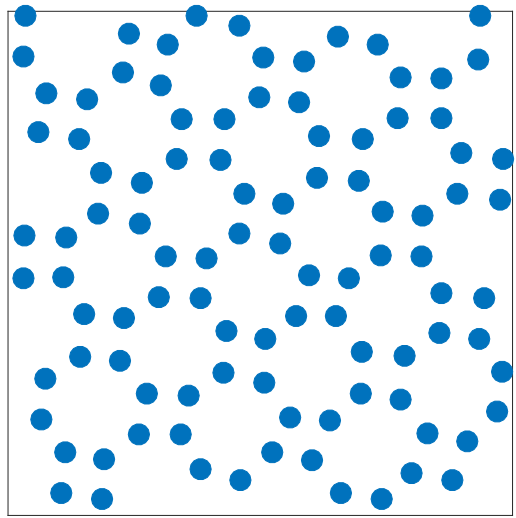}
	\caption{Monte Carlo simulation configuration for a system of particles interacting via the potential optimized for the TS lattice at $T=0.005$ and $\rho_0=1.03$.  As described in the text, this lattice self-assembled upon isochoric quenching to these conditions from a high temperature fluid.}
	\label{fig:truncsqu_quench}
	\end{figure} 
	
	As discussed in detail previously,\cite{SquLat_dmu_opt,KagSnub_gams_opt} to understand the stability of ground-state structures, it is helpful to consider the function $\psi(r)$
	       \begin{equation}
    	           \begin{aligned}
                        \psi(r) \equiv \frac{\phi(r)}{2}  - \frac{r \phi'(r)}{4}
               		\label{eq:psir}
               \end{aligned} 
            \end{equation}
  which determines the zero-temperature chemical potential $\mu_l$ of lattice $l$ via the relation $\mu_l=\sum_{i}^{r_{i,l} < r_c} n_{i,l}\psi(r_{i,l}(\rho_l))$, where $r_{i,l}$ denotes the $i^{\text{th}}$ coordination shell distance for that lattice at density $\rho_l$. In short, $\psi(r)$ quantifies the radially-varying `weights' (due to the form of the pair potential) that multiply the occupation numbers $n_{i,l}$ in a given lattice $l$ to determine the coordination shell contributions to its chemical potential. 
	\begin{figure}[ht]
	\begin{center}
		\includegraphics[trim=2cm 1cm 1cm 1cm,clip,scale=0.28]{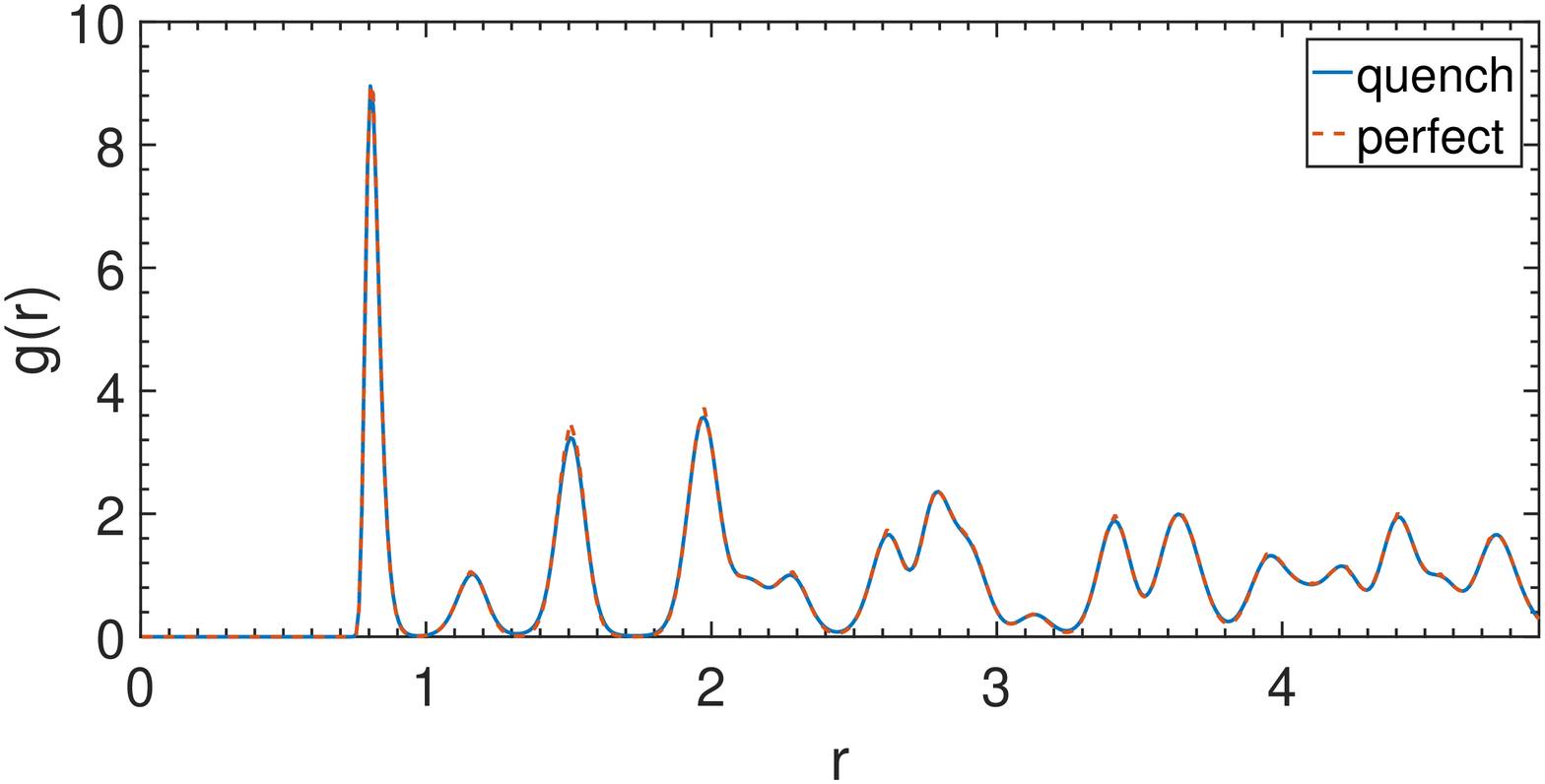}
		\caption{Radial distribution function $g(r)$ for the TS lattice at $T=0.005$ and $\rho_0=1.03$: (solid blue line) assembled via quenching from a high temperature fluid and (red dash line) equilibrated starting from the perfect lattice configuration.}
	\label{fig:truncsqu_gr}
	\end{center}
	\end{figure} 
	
  A plot of $\psi(r)$ is shown in figure \ref{fig:truncsqu_potpsi}b with vertical black lines corresponding to the first nine coordination shell positions of the TS crystal at the midpoint of its stable density range $\rho=1.03$. As seen, $\psi(r)$ displays two characteristic  plateau features: the first for separations in the range $0.7 \lesssim r \lesssim 1.3$ and the second for $2.4 \lesssim r \lesssim 2.7$. The function of these plateaus can be qualitatively understood as follows. The first plateau helps to destabilize standard Bravais and non-Bravais lattices (e.g., hexagonal and snub square patterns) which have relatively high coordination numbers (six and five in the first shell, respectively)--and, hence, higher contributions to the chemical potential--at these distances. The second plateau helps destabilize more closely related competitors that otherwise share or closely track the coordination shells of the TS lattice. For instance, the seventh shell of the target TS lattice is positioned right at the point where the second plateau starts to decrease ($r\sim2.6$) so that related shells for many of the stripe competitors at slightly smaller separations are destabilized more harshly. Lastly, the strongly repulsive `core' serves to destabilize competitors whose first shell is at a shorter distance than that of the TS lattice. Despite these features, note that the resulting $\psi(r)$ is still relatively smoothly varying, which--as discussed previously\cite{SquLat_dmu_opt,KagSnub_gams_opt}--is consistent with a target designed to display stability over a broad density range.

	Carrying out Monte Carlo simulations of particles interacting via the optimized pair potential as described in section \ref{sec:monte_carlo}, we verify the TS crystal can indeed readily assemble from the fluid phase upon isochoric cooling. A representative configuration of the assembled structure is displayed in figure~\ref{fig:truncsqu_quench}, showing that--aside from the usual minor defects due to the misalignment of the crystal and the boundaries of the periodically replicated simulation cell--a near defect-free TS lattice is obtained. The quality of the assembly is characterized more systematically (see figure \ref{fig:truncsqu_gr}) by comparing the radial distribution function $g(r)$ at the final temperature of the quench to that of an equilibrated crystal initiated from the perfect configuration at that temperature. As can be observed, $g(r)$ of the assembled system matches well with that of the equilibrium crystal. Note in particular the well resolved second peak, a shell where just a single neighbor is expected to reside. The fact that the assembled structure accurately captures it highlights the robustness of the optimized interactions.    

	\begin{figure*}[ht]
	\includegraphics[scale=0.30]{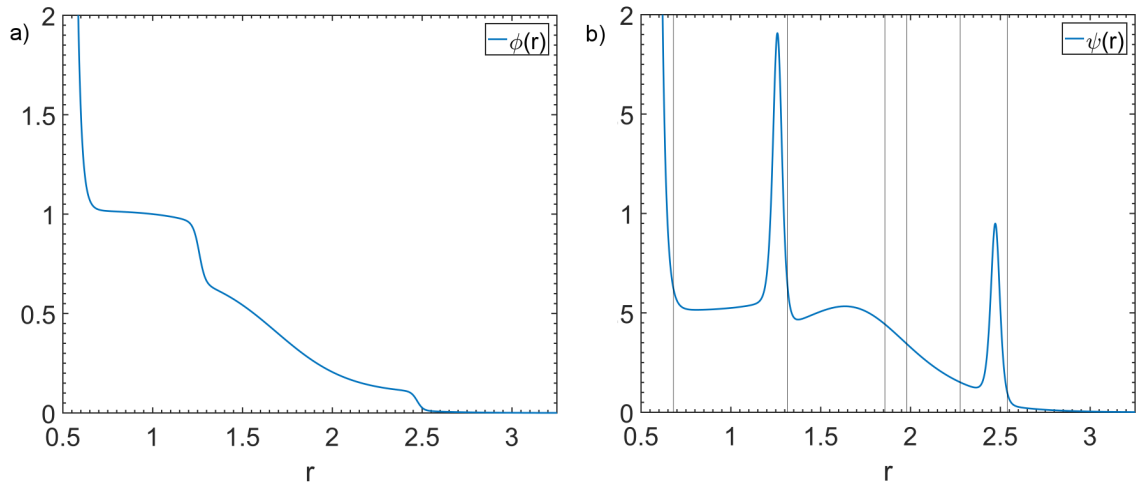}
	\caption{a) Repulsive pair potential $\phi(r)$ designed to stabilize the TH lattice as the ground-state structure, and (b) $\psi(r)$ obtained from $\phi(r)$ via eq.~\ref{eq:psir}. Black vertical lines indicate positions of the first six coordination shells of the TH crystal at the midpoint of its stable density range ($\rho=1.075$). The parameters of the optimized pair potential are presented in table S2.}
	\label{fig:trunchex_potpsi}
	\end{figure*} 

	The second target structure considered in this study, the TH crystal, provided a significantly more difficult design challenge. Despite the fact that the underlying structural motif of the TH lattice is similar to that of the TS lattice (see discussion below), we found that solution of the design problem for the more open TH lattice required consideration of nearly $50\%$ more competing structures as well as a more flexible pair potential (i.e., inclusion of a third hyperbolic tangent term in eq \ref{eq:potform}). While the pair potential $\phi(r)$ obtained from the optimization indeed stabilizes the TH crystal ground state, it does so only over a very narrow density range ($1.07 \leq \rho \leq 1.08$) and by assuming a more complex repulsive form (see figure \ref{fig:trunchex_potpsi}a and the associated parameters in table S2).  Note for instance the presence of two step-like features in $\phi(r)$ that are superimposed on a ramp-like repulsion similar to that of the optimized pair potential for the TS lattice. As shown in figure \ref{fig:trunchex_potpsi}b, this form gives rise to two sharp peaks in $\psi(r)$ at $r\sim 1.2$ and $r\sim 2.4$, which border a plateau region from $0.7 \lesssim r \lesssim 1.15$ and a broad hump from $1.35 \lesssim r \lesssim 2.35$. Each of these features are important for stabilizing the TH lattice relative to its competitors and can be understood as follows.  
	
	An analysis of coordination shell distances and occupation numbers shows that the plateau and broad hump features in $\psi(r)$ destabilize standard Bravais and non-Bravais competitors (hexagonal, snub square, honeycomb, etc) relative to the TH lattice because the former have more highly coordinated shells at those distances, and thus larger associated contributions to the chemical potential. It also shows that the sharp peak in $\psi(r)$ at $r\sim1.2$ destabilizes stripe classes a-c) (refer to figure \ref{fig:competitors}), which have first and second shell separations that closely track, but are slightly less than, those of the TH lattice. This is especially true for class c) stripes that have triangular motifs similar to those in the target structure. The main role of the sharp peak in $\psi(r)$ at $r\sim2.4$ is to penalize class d) and f) competitors whose first few shells share the same `Y' shaped motif  with the TH lattice (effectively shadowing TH shell distances) and thus can only be explicitly destabilized at these larger distances (more distant shells) where they display their stripe character. Lastly, note that the strongly repulsive `core' acts as an extra destabilizing factor for stripe competitors with first shells that are slightly closer in than those of the TH lattice.
	
	\begin{figure}[ht]
	\includegraphics[scale=0.40]{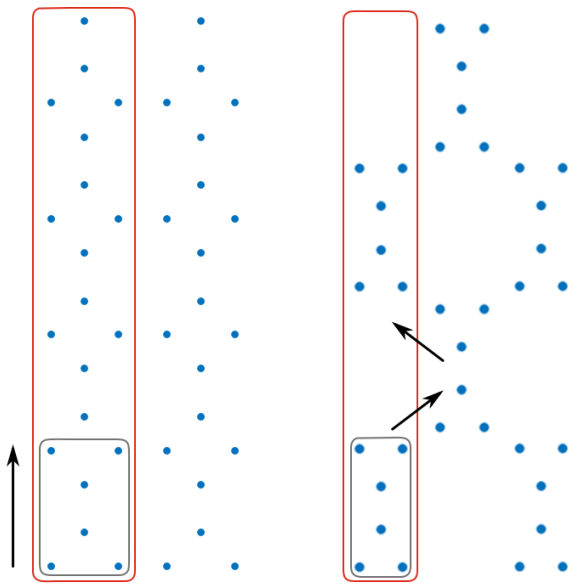}
	\caption{Design targets in the scope of stripe structures.  While TS and TH lattices have similar motifs (bottom gray rectangles) only the TS lattice can be cast as parallel stripes (left red box) spanned by the underlying motif. A similar approach to forming the TH lattice (right red rectangle) leaves out spaces in the stripe as per the staggered arrangement of the TH motif.}
	\label{fig:truncs_stripes}
	\end{figure} 
	
	To further understand why the TH lattice presents such significant design challenges not encountered for the TS lattice, consider figure \ref{fig:truncs_stripes}. Whereas the TS lattice (left) can be considered a class d) stripe structure (red rectangles) spanned by its internal motif (gray rectangle) with a specific inter-stripe distance, the TH lattice (right) cannot. Instead, the TH lattice displays a `staggered' arrangement of the internal motif. Translated into our design process, this means that while the TS lattice must only be stabilized against deformations of its motif and interdistance stripe configuration, the TH lattice must instead compete with whole classes of highly variable stripe configurations that mimic its underlying motif structure and make ring closure--the staggered configuration--difficult to realize.  This means narrow distinctions amongst many very closely related competitors that can only be meaningfully destabilized by sharply varying interactions (and the corresponding peaks seen in $\psi(r)$) that greatly complicate the optimization process. Consistent with this, the only other pair potential designed to stabilize the TH lattice\cite{RelEntropy_2D_structures} also exhibits such step-like features.   
	\begin{figure*}[ht]
	\includegraphics[scale=0.4]{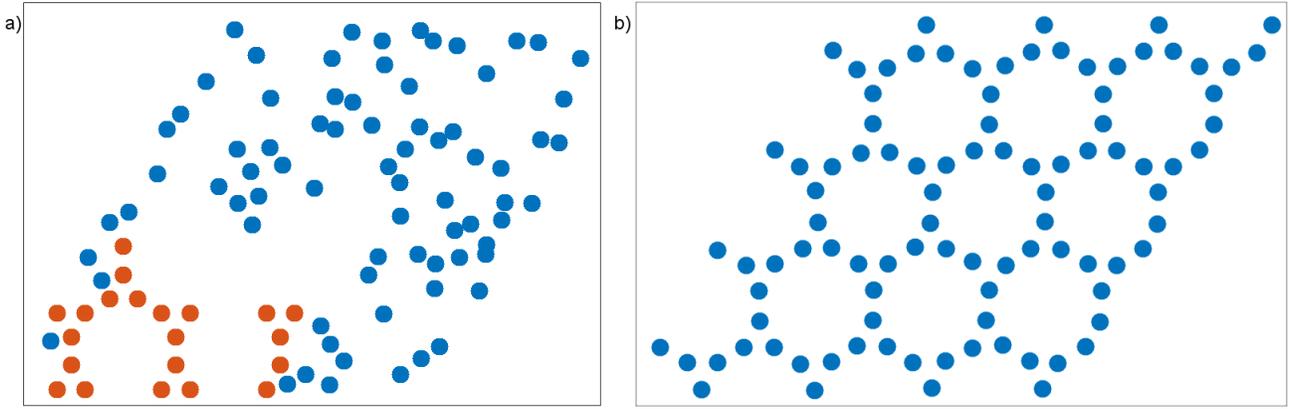}
	\caption{a) Initial configuration of a high temperature fluid at $\rho=1.075$, seeded with a small frozen TH crystal. Periodic boundary image chosen such that a complete seed is visible at bottom left section of the simulation cell. b) Configuration of assembled crystal after quenching to $T=0.06$ and equilibrating as described in the text.}
	\label{fig:trunchex_quench}
	\end{figure*} 

\begin{figure}[ht]
	\includegraphics[trim=1.5cm 0.5cm 1cm 0.5cm,clip,scale=0.28]{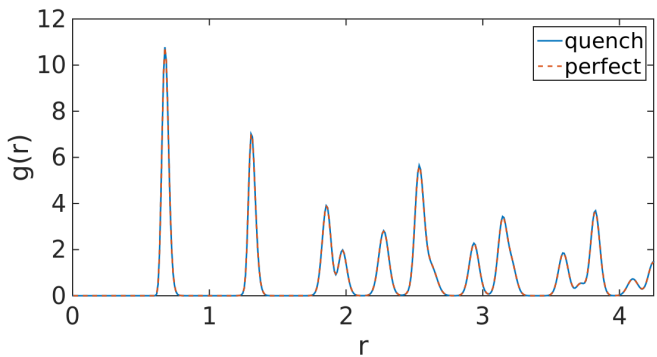}
	\caption{Radial distribution function $g(r)$ for the TH lattice at $T=0.06$ and $\rho_0=1.075$: (solid blue line) assembled via quenching from a high temperature fluid and (red dash line) equilibrated starting from the perfect lattice configuration.}
	\label{fig:trunchex_gr}
	\end{figure}

	Design challenges aside, we were able to verify self-assembly of the TH lattice from fluid configurations of particles interacting with the optimized pair potential via isochoric Monte Carlo temperature quenches. In this case, as described in section~\ref{sec:monte_carlo}, assembly of the target structure (on computational time scales readily accessible via simulation) required the addition of a small seed crystal during the quenching process.	As expected, success of crystallization depended largely on simulation time, with larger crystals or longer runs resulting in higher crystallization yield.  For results shown here, we used a seed size (21 particles) such that approximately 50$\%$ of parallel runs quenched into the crystal structure during the course of the simulation (see figure S3 for illustrative results). Shown in figure \ref{fig:trunchex_quench} are the initial and final configurations of one such seed run. The radial distribution of the assembled structure is provided in fig \ref{fig:trunchex_gr} and compared to that of a similar run started from the perfect crystal configuration at the final temperature and density. The excellent agreement shown demonstrates the success of the designed interaction for stabilizing the TH lattice.
\section{Conclusion}
\label{Conclusion}
	Using an efficient, recently introduced formulation of the inverse design problem for discovering interactions that favor a targeted ground-state crystal, we were able to determine repulsive, new isotropic interactions that stabilize open 2D TS and TH crystal lattices, respectively.  For the TS crystal, the optimized interactions stabilized the target structure in the ground state over a wide range of density, and particles interacting via the designed potential were shown to readily self-assemble into the TS crystal in isochoric Monte Carlo temperature quenches from a high-temperature fluid. 
	
	The open TH crystal proved to be a far more challenging design target, and its solution required consideration of significantly more competing structures as well as a more flexible repulsive pair potential. We demonstrated that while the TS crystal can be interpreted as a specific example of a stripe microphase, the TH crystal requires comparison against a highly varied field of stripe microphase competitors, and that the ring closure for the TH lattice required explicit staggering of underlying motifs that demanded very specific, sharply targeted interactions that greatly elevated the complexity of the problem. Despite this added difficulty, we found that particles with the designed interactions self-assemble into the TH crystal in isochoric Monte Carlo temperature quenches from a high-temperature fluid seeded with a small target crystal. 
 
\section{Supplementary Material}
	See supplementary material for figures of specialized TH competitors, optimized pair potential parameters and seeded TH Monte Carlo runs.  

\begin{acknowledgments}
T.M.T. acknowledges support of the Welch Foundation (F-1696) and the National Science Foundation (CBET-1403768). We also acknowledge the Texas Advanced Computing Center (TACC) at the University of Texas at Austin for providing computing resources used to obtain results presented in this paper.
\end{acknowledgments}


\begin{thebibliography}{32}
\bibitem [{Bra()}]{BravaisNote}%
  \BibitemOpen
  \href@noop {} {}\bibinfo {note} {{ While square and hexagonal could be said
  to belong to class a stripes as per our chart, these standard lattices are
  sufficiently common and important to be listed separately by name for clarity
  }}\BibitemShut {NoStop}%
\end{thebibliography}
%
\end{document}